\documentclass[preprint]{elsarticle}

\usepackage{amsmath,amssymb,amsfonts}
\usepackage{algorithmic}
\usepackage{graphicx}
\usepackage{textcomp}
\usepackage{xcolor}
\usepackage{balance}
\usepackage{blindtext}
\usepackage{url}
\def\BibTeX{{\rm B\kern-.05em{\sc i\kern-.025em b}\kern-.08em
    T\kern-.1667em\lower.7ex\hbox{E}\kern-.125emX}}

\begin{document}

\title{Enabling Machine Learning-Ready HPC Ensembles with Merlin}

\author{J. Luc Peterson\corref{cor1}}
\ead{peterson76@llnl.gov}

\author{Ben Bay}
\author{Joe Koning}
\author{Peter Robinson}
\author{Jessica Semler}
\author{Jeremy White}
\author{Rushil Anirudh}
\author{Kevin Athey}
\author{Peer-Timo Bremer}
\author{Francesco Di Natale}
\author{David Fox}
\author{Jim A. Gaffney}
\author{Sam A. Jacobs}
\author{Bhavya Kailkhura}
\author{Bogdan Kustowski}
\author{Steven Langer}
\author{Brian Spears}
\author{Jayaraman Thiagarajan}
\author{Brian Van Essen}
\author{Jae-Seung Yeom}

\cortext[cor1]{Corresponding author}
\address{Lawrence Livermore National Laboratory, Livermore, California 94550, USA}

\newcommand{\merlin}{\texttt{\small Merlin}}
\newcommand{\epi}{\texttt{\small epicast}}

\begin{abstract}
With the growing complexity of computational and experimental facilities,
many scientific researchers are turning to machine learning (ML)
techniques to analyze large scale ensemble data. With complexities
such as multi-component workflows, heterogeneous machine architectures,
parallel file systems, and batch scheduling, care must be taken to facilitate
this analysis in a high performance computing (HPC) environment. In this
paper, we present \merlin, a workflow framework to enable large ML-friendly
ensembles of scientific HPC simulations. By augmenting traditional
HPC with distributed compute technologies, \merlin~aims to lower the
barrier for scientific subject matter experts to incorporate ML
into their analysis. In addition to its design, we describe some example applications that
\merlin~has enabled on leadership-class HPC resources, such as
the ML-augmented optimization of nuclear fusion experiments and the calibration of 
infectious disease models to study the progression of and possible mitigation strategies for
COVID-19.
\end{abstract}

\maketitle

\begin{keyword}
workflow management \sep sampling \sep simulation \sep machine learning
\end{keyword}

\section{Introduction}
\label{sec:intro}
\subsection{The Problem: Using Machine Learning to Bridge Simulation
and Experimentation}
A fundamental tenet of scientific investigation is the challenging of
models with experimentation.
A process that began as a search for analytic descriptions of reality has
evolved into the validation of complex computational calculations with
increasingly complicated experimental observations. Large scale
experiments, such as those performed at the National Ignition Facility (NIF)~\cite{mosesNIF}, 
can produce upwards of a gigabyte of multi-component information on each shot.
Furthermore, modern high performance computing (HPC) platforms
have made it easier than ever to produce large quantities of simulated
data. For example, a single instance of a moderately-resolved radiation
hydrodynamic simulation of a NIF experiment can produce of order $10$ gigabytes of simulation data 
in six hours on a single Intel Haswell node~\cite{peterson-zonalflow}.

In addition to being able to produce large quantities of data from a single simulation and motivated in part by the emergence of scientific machine learning (ML), the creation of large-scale scientific simulation databases is an emerging trend in
HPC. 
Simulations can provide multivariate data (e.g. 
time-series, scalar, vector, and image) that are physically correlated
\textit{by design}. Due to their complexity and scale, large simulation datasets 
with large numbers of samples/instances are an attractive challenge 
to ML systems. Of the publicly 
available datasets in the UCI ML repository~\cite{Dua:2019}, the largest
sample size consists of 63 million instances of time-series medical
sensor data~\cite{Schmidt:2018uq}. Of the 49 physical science datasets, 
the largest consists of roughly 11 million particle physics 
simulations~\cite{Baldi:2014yq} intended for scalar classification. These datasets mainly 
contain scalar numbers (either static or as time-series), so they are gigabytes in size. 

Ensemble simulations that produce data on a grid or generate synthetic images may  
produce much larger numbers of samples and much larger datasets. 
An ensemble of 10,240 simulations of Earth's weather was run on the 
K Computer~\cite{AtmosEnsemble}. This ensemble produced 1.21 TB of data and 
roughly 3.3 million grid slices. A grid slice can be thought of as a 40,960 ``pixel" 
image (i.e. one pixel per grid zone). As a benchmark for HPC-based simulations of 
multivariate data (scalar, vector, image), the authors of \cite{peterson-zonalflow} used 
an on-node producer-consumer task queue system~\cite{langer2016hydra} to 
create a 70 TB database of 60,000 multiphysics simulations. The dataset includes 
roughly a billion compressed synthetic x-ray images as well as time series and scalars.

As our capacity to produce scientific data has increased, our
ability to analyze it has largely failed to keep up. A typical comparison
between a NIF experiment and simulation involves the distillation of all
data into around 10 scalar quantities, a process that leaves most of the
data unanalyzed and the models potentially under-constrained.

One hope is that modern data analytics and machine learning will
provide the tools to help scientific analysis keep up with scientific data
production. Possible applications include performing near real-time feature selection for 
data compression or creating surrogate models for expensive simulation 
codes. Surrogate models could be used to perform  
sensitivity studies, optimization, uncertainty quantification, 
and validation of computer models against experiments without running additional HPC simulations. 

ML systems require many examples (samples) to build accurate surrogate 
models, but HPC systems are designed to execute at most only a few concurrent instances of
very complex models (e.g. large MPI jobs). This underlying tension between ML requiring many
simulations but HPC being optimized for a few means that the creation of HPC
simulation datasets used to train ML models must be 
accomplished with care. Standard HPC workflow toolkits may not be the most 
efficient way to produce the large ensembles of simulation data required for ML 
model training.
 
In this paper, we detail the development of a simulation
workflow framework, \merlin, whose purpose is to facilitate the creation of
large scale ensembles of simulation data suitable for analysis by
machine learning tools.

\subsection{Challenges for ML Ensembles in HPC}
\label{sec:ml_challenge}
As a motivating example for the application of ML to an HPC environment, 
consider a simple ``intelligent" sampling workflow,
which uses active learning to dynamically select new simulations to
add to an ensemble. Since HPC simulations are expensive, it is reasonable 
to only run simulations if necessary to improve the accuracy of the trained 
model. The simplest loop to accomplish this involves running simulations, 
post-processing the raw data from those simulations into salient features, 
training an ML system on those features, evaluating the error on that 
training and adding new simulations to reduce the model error.

A major challenge of even this simple workflow is that it is
inherently heterogeneous, involving multiple simulation codes (at the
least, a simulator and a learner). The simulator must calculate the raw
quantities of interest and write them in a data format that can be read
efficiently by the learner. However, the simulator and learner 
are likely optimized for a different set of execution parameters. For
instance, learning can be extremely
efficient on a graphics processing unit (GPU), but simulating might not
be, depending on the application. If the raw data produced by the
simulator need to be post-processed prior to ingestion by the learner,
such a post-processor would likely have different requirements. This
heterogeneous workflow must be coordinated and orchestrated in an
efficient manner, potentially across multiple batch allocations and 
hardware systems, which may or may not be able to directly access each other.

Modern HPC systems were designed and optimized for the efficient
execution of a few large scale simulations, instead of the large number of
smaller scale simulations necessary to train accurate ML models.
Batch scheduling systems are generally not designed to launch thousands to millions of
simulations. Parallel file systems can lose performance when presented
with large numbers of concurrent reads and writes that overwhelm
meta-data servers. Dynamically loaded
shared objects can present similar problems.

Fundamentally, creating ML-ready ensembles of HPC simulation data necessitates a
workflow technology that can efficiently coordinate asynchronous heterogeneous
simulation tasks at scales well beyond the designed operation of HPC systems. 
\merlin~aims to fill this space.

\subsection{Related Existing Workflow Technologies}
Workflow technology is a rich field, where many
approaches provide similar functionality. Their
underlying mechanisms, implementations, and flexibility vary widely. These
technologies range from full workflow tools to scripting languages for
procedurally generating tasks. In this section, we briefly
survey some existing workflow technologies that have been applied to scientific
computing.

Pegasus~\cite{deelman-fgcs-2015} and Fireworks~\cite{CPE:CPE3505} are  
full workflow systems which provide task tracking, generalized scheduling, 
and Python APIs for creating workflows programmatically. Both allow
tasks to be defined using directed acyclic graphs (DAGs). The
Fireworks ``Launchpad" (built on MongoDB) is a staging mechanism
which pushes bundles of work to compute nodes. Pegasus 
makes use of HTCondor through DAGMan and follows a pull model.

The Sandia Analysis Workbench (SAW) provides users a GUI
interface and plugin system using the Eclipse IDE and their integrated OSGi
plugin system~\cite{friedman2015incorporating}. 
SAW provides flexibility by implementing plugins as generalized extension points 
representing abstract configurations that define an interface. Plugins
can be tied together to build larger workflow processes, in particular
workflows geared towards verification and validation testing (V\&V). All
components implement a backend Java interface to their workflow engine,
allowing components to be used to create an explicit workflow graph and
enables task scheduling and monitoring.

The UQ Pipeline~\cite{dahlgrenscaling} has application
specific capabilities such as performing sensitivity studies, estimating
parameter values, and a host of other features. The UQ Pipeline makes use of
CRAM~\cite{Gyllenhaal:2014et} to consolidate jobs in a batch system so that they can 
achieve high sample
counts in their workflows without submitting huge numbers of jobs to HPC clusters.

On the other end of the spectrum, Swift~\cite{Wilde:2011:SLD:2286659.2286714}
is a popular parallel scripting language. It supports most of the structures
expected from a programming language and can create
dynamic workflows. Like Pegasus, it follows a data flow model. Swift 
follows a functional approach, where functions take in data inputs and
produce data to be passed to other functions.

While these technologies provide a robust set of features, they are not without their 
challenges. In broad terms, these workflow technologies fall under two categories: 
distributed lightweight computing or localized large scale HPC.

Distributed workflow systems tend to have larger sets of features, but provide domain 
scientists, who are used to working in HPC environments,
with an expensive upfront training cost, and can potentially lock scientists into release 
cycles out of their control. The lack of control on new features can be limiting,
especially as new forms of compute hardware become available (GPUs, ASICs,
etc.) and when testing on leadership class machines. Security imposes additional 
constraints in
HPC environments, and retaining system security and integrity often disqualifies 
commercial tools
due to the necessary efforts to verify software before use. Other limitations
include the necessary traceability of user actions which disqualifies services
that obfuscate such information. More fully
fledged workflow tools like Pegasus and Fireworks are attractive, but obtaining the 
security approvals to stand up their backend technologies
in HPC environments can prove cumbersome.

HPC-focused workflow technologies, such as the UQ Pipeline, Swift, 
CONDOR~\cite{Foster:2017fk} and EMEWS~\cite{Ozik:2016wm} (the latter two built 
upon Swift), focus on executing many tasks in a single large MPI job running on a 
single machine. Approaches like Swift that avoid coordination via the filesystem 
show advantages in scaling, but their focus on single batch job ensembles precludes cross-
system workflow coordination. On the other hand, Maestro~\cite{di_natale_maestro_2019} 
can coordinate across batch jobs, but does so via filesystem coordination and live 
background processes running on login nodes, limiting throughput. Programmatic interfaces, 
such as those in the UQ Pipeline and Swift, and GUI-interfaces like in SAW, represent 
barriers for user adoption, since their interfaces represent a new language for subject matter 
experts to learn. This contrasts with Maestro's YAML 
specification for DAG definition, which has users define workflow steps in shell syntax.

\subsection{Summary of \merlin~and Paper Outline}
In this paper we explore the creation of a lightweight, scalable workflow
suite specifically designed to enable large ensembles of HPC simulations
tailored for easy analysis by machine learning tools. The software, \merlin,
combines workflow and HPC technologies in a framework that
enables the scalable execution of heterogeneous workflows suitable for next-generation
ML-driven workflows. It is compatible with present day HPC
systems, security models and front-line
multiphysics codes. Sec.~\ref{sec:merlin} details the design of \merlin,
while Sec.~\ref{sec:applications} demonstrates its use on a variety of
applications. In particular we show that \merlin~can handle a variety
of ensemble types, from the efficient execution of many lightweight tasks,
to multi-machine heterogeneous simulator-learner optimization workflows to
large multiphysics-based ensembles. We demonstrate \merlin's
capability by reporting on the creation of an unprecedentedly large inertial confinement fusion (ICF)
dataset, consisting of approximately 100
million simulations and 4.8 billion images, that was generated on the Sierra
Supercomputer at Lawrence Livermore National Laboratory. Finally,
we show that \merlin 's ability to apply surge computing resources to complicated dynamic workflows
helped enable the quick re-calibration of an existing infectious disease model to study the progression of
and possible mitigation strategies for COVID-19.

\section{\merlin}
\label{sec:merlin}
An ML-integrated workflow system designed to create
massive ensembles of multi-physics simulations has a number of design
considerations that are not present in smaller workflows. In particular, such a system must be able to:
\begin{itemize}
    \item Operate in an HPC environment
    \item Produce and process the data from $>10^5$
        MPI-driven simulations
    \item Support both in-situ (on-node) and in-transit (concurrent) analysis
    \item Support multi-machine and multi-batch slot workflows
    \item Support multiple executable types
\end{itemize}
With these considerations as motivation, we next discuss the
technological choices that went into \merlin~and perform a series of tests on its performance. 

\subsection{Design Considerations}
The primary design consideration for \merlin~is the
environment in which large scale simulations must be run, namely large
HPC systems such as Sierra. These systems operate at the leading
edge of computational technology and at a uniquely large scale, under
constrained access models. Efficiently executing a single instance of a
simulation at these scales can require customized libraries, compilers,
environment setup, and control. They also require effective interfacing with a system
batch scheduler, which may, along with the operating system on the
machine, be unique. At these large scales, failures of hardware and
software are to be expected.

These constraints alone imply that a ``one size fits all"
approach to a workflow is likely insufficient. Put another way, subject
matter experts need to be able to have fine and programmatic control over the
execution of their component applications. Individual
components may need to run on different machines, each with their own
batch scheduling systems and potentially their own isolated file systems.
Furthermore, the competition for available computational resources on
these machines can be fierce, with simulation throughput highly
dependent on the background load on the system.

Beyond the unique operating environment of heterogeneous
state-of-the-art HPC systems, ensembles of multi-physics simulations will
likely need to consist of a large quantity of individual samples. As an
example, the machine learning model used in
\cite{peterson-zonalflow} was trained on an ensemble of roughly
$60,000$ ICF simulations. Since ML-systems require a large number of
individual sample points to obtain an accurate representation of the
response surface, a suitable workflow needs to be able to accommodate
and coordinate the execution of at least tens of thousands of MPI-based
simulations.

Another consideration is that these multi-physics codes can produce a
large quantity of raw data, for instance mesh-based fields such as
temperature and density. Prior to being fed into a machine learning model,
these raw quantities must be post-processed into derived quantities, such
as synthetic diagnostic signatures, or hand-curated features. While the
raw data itself can be quite large, of order gigabytes or more per
simulation, the derived quantities themselves can be of
order megabytes, depending
on the type of data of interest. This necessity to post-process implies that
an ML-driven workflow needs to be able to handle both \textit{in-situ}
(processed in-line by the parent simulation code) and \textit{in-transit}
(processed by a separate data processing step) analysis.

Additionally, an ML-driven large scale HPC ensemble
may need to be executed on
multiple machines across multiple batch slots.
The need for multiple batch
slots is straightforward - many leadership class
machines have time limits to the
length of a single job. Unless all simulations in an ensemble can
complete on the allocated resources prior to the time limit, the workflow
will need to be able to coordinate and spill-over into multiple resource
allocations. These batch slots, however, might not be on the same
machine. For example, the ML systems might train most efficiently on
different hardware than is optimal for the main simulation code. 
Such an ensemble would consist of different executables, whose work
needs to be coordinated.

All of these considerations motivate a series of technological
choices to more easily enable ML-suitable large ensembles. Scalability
arguments suggest that simulation coordination needs to be
accomplished
via message passing, instead of queries to files or the batch system. The
need for multiple machines and batch systems implies that workflow
coordination should persist outside of any particular job. The unique HPC
environments and the customization required to effectively execute multiphysics
codes on leadership class machines means that the workflow interface needs to
accommodate the shell-based commands subject matter experts require for their
parallel jobs. Data formats should be portable across machines and languages,
with the capability for in-memory and on-disk inter-operability. ML-driven ensembles may 
require distributed and dynamic task creation to avoid
bottlenecks at scale and to enable intelligent sampling schemes.

\subsection{System Design}
We describe in this section  \merlin, its underlying software components, its user-facing interface, and its task-generation algorithm. The primary requirements set out in the previous section are scalability, flexibility, and ease of use.

To enable cross-machine and dynamic task creation, \merlin~is built upon a producer-consumer workflow model. As an underlying library, we choose Celery~\cite{celery} for both its asynchronous task scheduling and its ability to rely on multiple task brokers and results backend. While \merlin~can be configured for any of Celery's supported brokers and backends, defaults are RabbitMQ~\cite{rostanski2014evaluation} and Redis~\cite{han2011survey} for their scalability and reliability.

As an interface to Celery, we leverage Maestro~\cite{di_natale_maestro_2019}. Maestro has a shell-like interface for defining workflows. A user defines a ``study" with distinct ``steps" of different ``parameters" and dependencies. The dependencies and parameters define a directed acyclic graph (or DAG). Each step has a command section written in an arbitrary shell language\footnote{We extend Maestro's original syntax by setting the shell to be a user-defined parameter that can vary with the step. This allows for steps to be written not only in        different shells (like bash or tcsh), but also more complete languages, such as python or lua. In addition to this step-wise syntax customization, the use of an artibrary shell facilitates composite multi-language workflows.} with variable key words for parameters and DAG-workspaces. At runtime, the DAG gets expanded with the full set of parameters replacing the key words and the dependency chains completed. \merlin~takes these commands and dependencies and translates them into Celery task constructions of subprocess commands, which are executed by workers receiving the task in a directory unique to that task.

The shell-like interface allows flexibility in both workflow component syntax (for instance to call MPI-driven components through flux \cite{ahn2018flux}, to leverage libraries like Conduit~\cite{conduit_2019} for data exchange and translation or to execute components in containerized environments). When combined with the DAG meta-construct and a producer-consumer model, a wide variety of workflow topologies are supported. For instance, workflow steps can make additional calls to \merlin~to enqueue other workflow studies, or can use language syntax to implement dynamic flow control and branching. \merlin's use of Maestro as an interface to Celery allows for workflow flexibility in a programmable and HPC-intuitive syntax.

\begin{figure}[]
    \centering
    \includegraphics[width=\columnwidth]{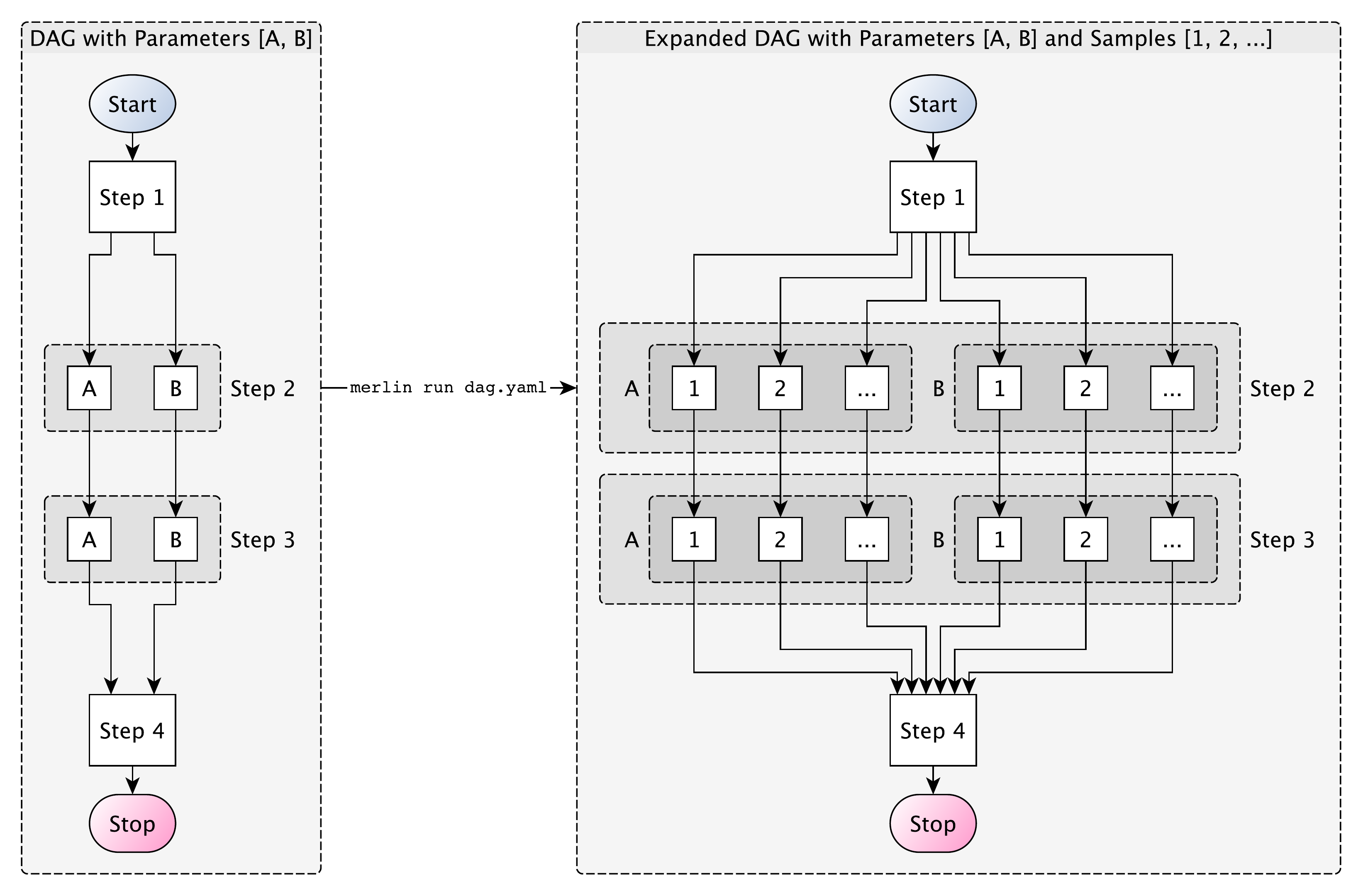}
    \caption{\merlin~takes a compact directed acyclic graph (DAG) with discrete parameter values (A, B, ...) and dependencies and expands it to include hierarchically and dynamically generated samples (1, 2, ...), which are layered onto the parameter set. Samples have a known topological dependence, which can be exploited for performance, whereas DAG dependencies can be more complicated and require more time to calculate. A layered approach allows for both flexibility and scalability.
     \label{fig:dag-samples}}
\end{figure}
For scalability, we augment Maestro's notion of workflow ``parameters" with ``samples." Samples are variable sets that are executed for each distinct parameter combination (see Figure~\ref{fig:dag-samples}). This makes their workflow topology simple (they are embarrassingly parallel for a given parameter combination), a fact that we exploit for scalable task generation and execution.

In particular, \merlin~takes advantage of the producer-consumer workflow model inherent to Celery to construct a hierarchical task generation algorithm. At producer runtime (via the command \texttt{merlin run}), the algorithm parses the total number of samples to be executed and breaks them into a hierarchical grouping of multiple levels. Instead of enqueuing the entire ensemble of samples at once, \merlin~enqueues only a single task at the top of the hierarchy containing the metadata necessary to enqueue its child tasks. The process recursively executes down the hierarchy until reaching the bottom nodes, at which point the actual simulations are enqueued. As consumers (\texttt{merlin run-workers}) come online, they begin to dynamically populate the task queue, in addition to beginning to work on actual simulations.

As an additional design feature, we exploit Celery's ability to set tasking priority and explictly prioritize simulation tasks over task-creation tasks. In testing, this significantly improved server stability for very large ensembles by reducing the hardware constraints on the server. Without such a feature it's possible for one user to enqueue a large number of tasks, effectively reserving space on the server for their workflow without consuming those tasks. This problem becomes particularly acute for real-world simulations with long execution times: task-creation is fast but task-consumption is slow, so task-creation quickly outpaces task-consumption and strains the server. However, by telling workers to prioritze draining the queue over filling the queue, we guard against this pathology. 

\begin{figure}[htbp]
    \centering
    \includegraphics[width=0.8\columnwidth]{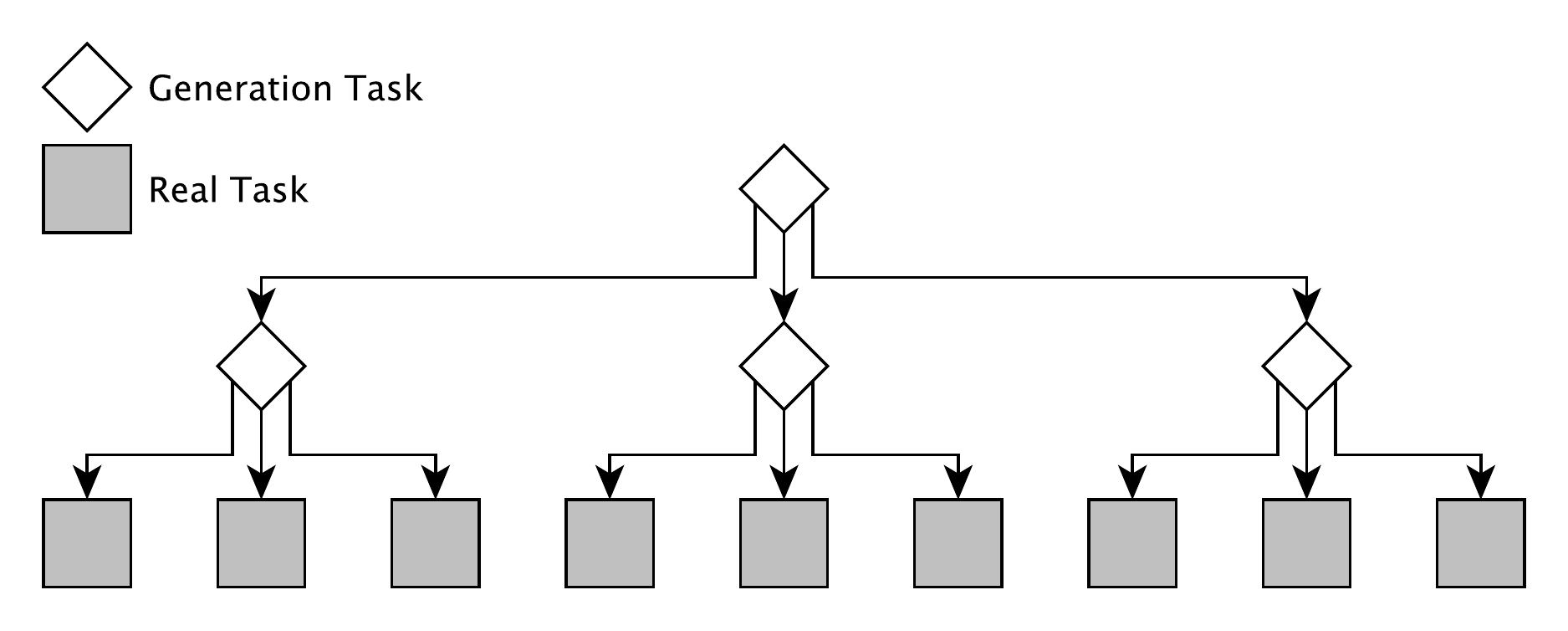}
    \caption{Total tasks in a simple hierarchy of 9 real tasks, divided with up to 3 tasks per level. Generation tasks (white diamonds) dynamically generate child tasks and populate the queue server. True workflow steps are executed in real tasks (gray squares).
     \label{fig:simple_h}}
\end{figure}

As an example, consider using 4 workers to execute the workflow displayed in Fig.~\ref{fig:simple_h}. It has 9 samples (``real tasks'') defined in a hierarchy of up to 3 branches at each level (for a total of 3 levels: a single node at the top, 3 in the middle layer and 9 on the bottom). Upon execution, a single task is enqueued and executed by the first worker, which creates 3 more tasks (the middle layer). Each of these are executed asynchronously by the other available 3 workers, which each create 3 tasks, the bottom of the tree, comprised of the actual simulations. As soon as the first simulation task in enqueued, the original first worker, having finished its generation tasks and now sitting idle, sees the new task and begins executing the first simulation. Meanwhile the other workers finish creating the other 8 simulation tasks. As each worker finishes its generation tasks, it can start working on a real task. In this fashion, the task queue gets dynamically populated and drained.

The task creation hierarchy provides a few benefits. Firstly it means that task enqueuing is an efficient process, since \texttt{merlin run} needs only to populate the queue server with the metadata required to create the tasks, not the tasks themselves. Secondly, it makes task creation a scalable process itself, since the load is spread out dynamically among workers. Finally, it means that actual simulation work can begin executing in a shorted wallclock time, as soon as the first simulation task is enqueued, instead of having to wait for the entire simulation ensemble to be created.

\subsection{Performance Analysis}
In this section, we perform a series of tests to measure the performance of \merlin~on an idealized but simple workflow of independent tasks each consisting of the shell command \texttt {sleep 1}, which represents a null simulation. To create unique samples the workflow step also includes a comment with the sample identification. The goal is to measure the overhead associated with using \merlin~to run simulations. We choose 1 second because real world HPC simulations are unlikely to complete more quickly and a sleep time of 0 seconds could allow system level jitter to confuse results. 

All of these tests were performed on the Pascal supercomputer at Lawrence Livermore National Laboratory, with access to a standalone in-house RabbitMQ and Redis server, residing on a single node of an adjacent machine in the same compute center and accessible by all compute nodes of Pascal.

\begin{figure}[htbp]
    \centering
    \includegraphics[width=\columnwidth]{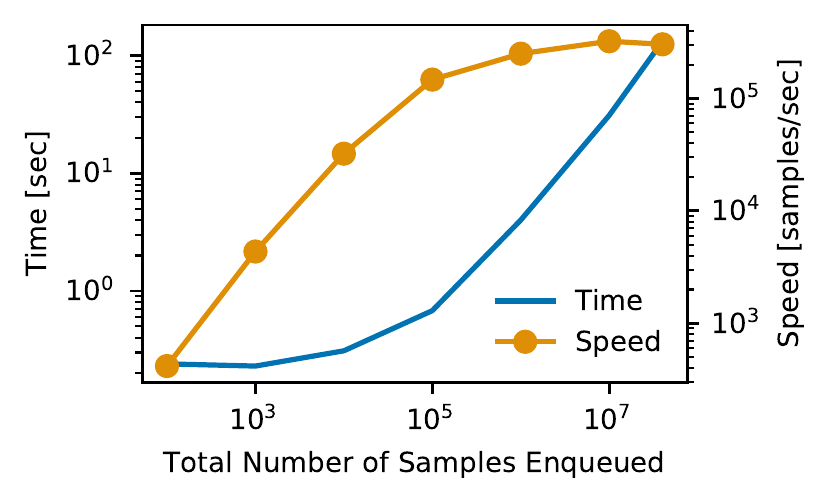}
    \caption{Scalability of task enqueuing time [seconds] and speed [samples/second] with total number of samples.
     \label{fig:enqueuing_time_and_speed}}
\end{figure}
The first test times how long it takes to enqueue studies of different sizes. Figure~\ref{fig:enqueuing_time_and_speed} measures the time it takes \texttt{merlin run}~to create task hierarchy metadata and populate the queue server, for simulation ensembles of various sizes. Both the total time required to populate the server as well as the enqueuing speed (defined by the ratio of the total time to the total number of samples) are plotted for studies ranging from 100 samples up to 40 million samples. Above 40 million, the RabbitMQ server hit a system-defined limit on allowable message sizes (2.1 GB). The task enqueuing reaches a peak speed around $3\times10^5$ samples per second and plateaus at ensemble sizes greater than $10^5$. At smaller ensembles task queuing speed is lower due to the overhead of creating the ensemble data structure dominating the process. Nonetheless, ensembles at this scale can be queued in less than 1 second, and 40 million simulations take around 100 seconds to queue.

\begin{figure}[htbp]
    \centering
    \includegraphics[width=\columnwidth]{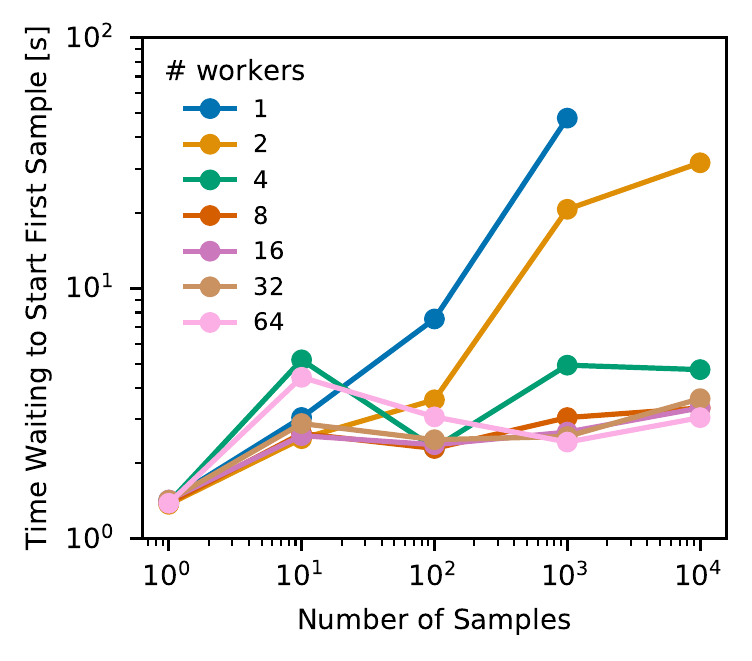}
    \caption{Pre-sample startup time. Number of seconds taken by a workflow between the activation of workers and the starting of sample processing.
    \label{fig:startup_time}}
\end{figure}
Figure~\ref{fig:startup_time} measures how long workers wait before executing the first simulation. This represents the time spent unpacking the task hierarchy before simulations begin. Minimizing this time means that batch allocations and their dedicated resources do not lie idle during ensemble initialization. As ensemble size increases, the time spent waiting to start simulations increases as well; however, adding extra worker threads causes this time to drop, for instance allowing a 1000-sample ensemble to begin in 3 seconds with 4 worker threads as opposed to 50 seconds with only 1 thread. Adding additional workers doesn't appreciably lower the time, since there are enough workers to unpack down to the first simulation leaf of the hierarchy.

Together, Figures~\ref{fig:enqueuing_time_and_speed} and \ref{fig:startup_time} demonstrate that the task hierarchy generation algorithm reduces the overhead associated with creating simulation ensembles. They make \merlin~an effective producer-consumer workflow model with a nearly non-blocking producer stage.

\begin{figure}[htbp]
    \centering
    \includegraphics[width=\columnwidth]{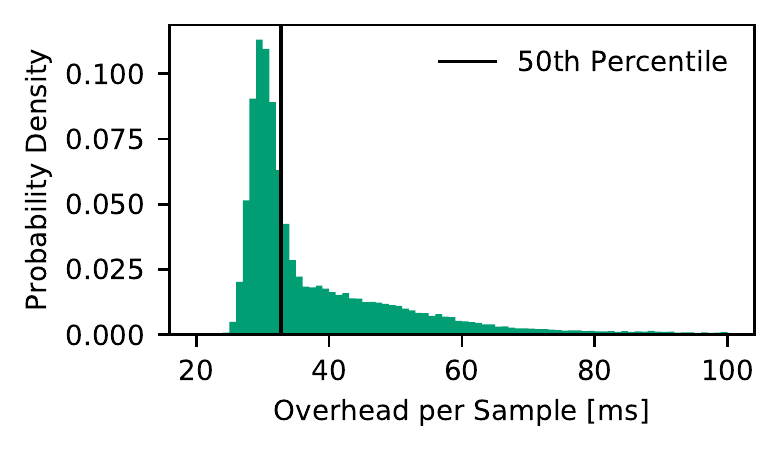}
    \caption{Histogram of all individual task times for approximately 900,000 tasks spawned by the 48 workflows comprising this overhead study. Not all tasks are equal; in the hierarchical task creation algorithm, there are different classes of tasks (task-creation, generic workflow steps). This heterogeneity plus system jitter likely result in the right-skewed distribution. Observations with a modified z-score greater than 5 are classified as outliers and excluded from the plot. The bulk of system tasks are small and take less than 50 milliseconds.
    \label{fig:histogram}}
\end{figure}
Another measure of overhead is the time spent by the workflow system not running the simulation itself. Figure~\ref{fig:histogram} shows a histogram of this overhead for all roughly 900,000 simulations run in the studies in Fig.~\ref{fig:startup_time}, as defined as the time between when a worker acknowledges receiving a task and when it tells the central RabbitMQ server it has finished, minus the 1-second sleep interval. This time includes the time needed to create a unique directory in which to execute the sample, create the script that contains the sample-unique instructions and any network-related traffic. Average overhead is tens of milliseconds per sample, with a median value of 32.8 ms. The mode peaks slightly below that, but a long tail extends out towards 100 ms. Nonetheless, the overhead associated with each sample is small compared to the expected compute time for an HPC-based simulation.

\begin{figure}[htbp]
    \centering
    \includegraphics[width=\columnwidth]{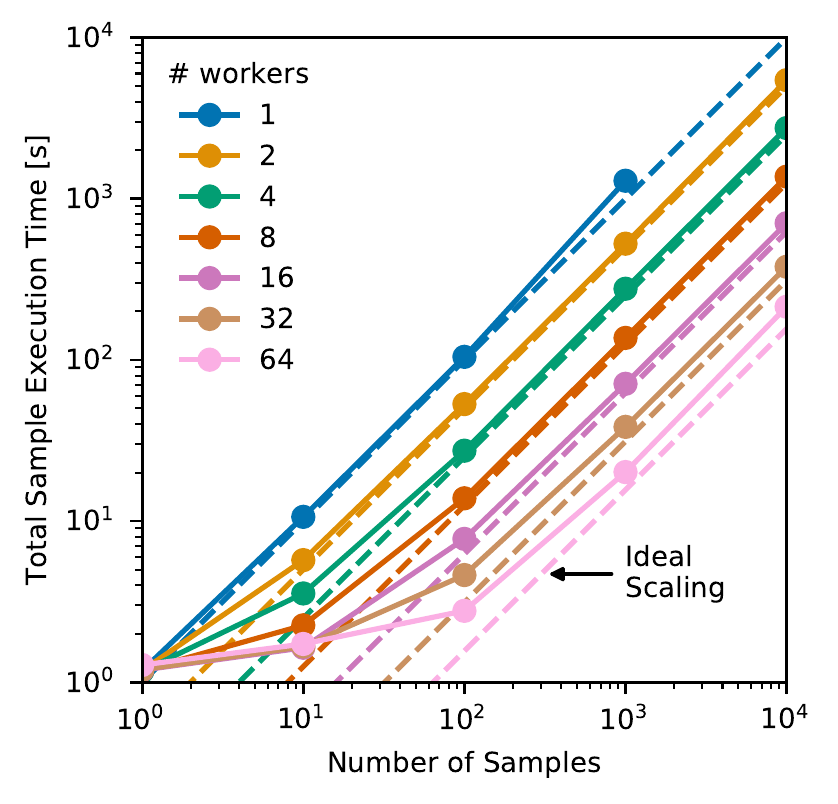}
    \caption{Sample task time. As more workers are added, heavy sample overhead is reduced. This is most easily observed at high sample volumes.
    \label{fig:sample_time}}
\end{figure}
Figure~\ref{fig:sample_time} shows the total time required to execute all of the simulations as a function of the number of available workers (solid with circles) alongside dashed curves of ideal scaling, which correspond to each worker taking exactly 1 second to execute the null simulation. In other words, we expect $10^3$ samples to take $10^3$ seconds with one worker thread but half the time with two workers. As the number of samples increases, the data trend towards the ideal scaling, showing that the overhead from ensemble generation and processing does not increase. Likewise, adding more workers has the expected scaling result of allowing the ensemble study to get executed more quickly: doubling the workers halves the time needed to process the samples.

Since the workers are decoupled from the work, Fig.~\ref{fig:sample_time} demonstrates that \merlin~workflows can effectively take advantage of surge compute resources (just-in-time and dynamic utilization of newly available resources). As more workers come online, they can connect to the central queue server and begin processing work alongside those that are already running, without adding additional workflow overhead.

In summation, a series of null 1-second tests of varying sized ensembles quantified the overhead associated with queuing tasks, unpacking ensembles and running simulations. This overhead is small compared to what is expected for HPC-based simulations. They also showed worker scaling performs as expected. \merlin's task hierarchy algorithm allows for large ensembles to be queued and unpacked in seconds. When combined with the natural decoupling of workers from work, \merlin~becomes an effective producer-consumer workflow paradigm for HPC-based simulation studies.

\section{Example Applications}
\label{sec:applications}
In this section, we deploy \merlin~on a few real world scientific examples. We focus not on the scientific results of the studies themselves, but rather on the workflow aspects of them, how they were constructed, how they used \merlin~to accomplish their goals, and what generic lessons can be learned from real world applications. Two examples are from physics, specifically inertial confinement fusion, and one is from biology, namely epidemiological studies of COVID-19 progression and mitigation. The first physics study uses JAG~\cite{gaffney2015data}, an analytic single core python-based simulation code, and leverages the Sierra supercomputer to scale out this simple model and produce a large physics-based dataset for machine learning applications. The second physics study uses the MPI-driven multiphysics code HYDRA~\cite{HYDRA-ref} in a machine-learning augmented optimization loop of a fusion experimental design. The COVID-19 study is a nested two-phase workflow that first calibrates an MPI agent-based epidemiological model \epi~\cite{epicast} and then uses the calibrated model to simulate various intervention scenarios for multiple metropolitan areas simultaneously. This set of studies collectively covers large-scale, multi-component, ML-augmented, iterative, and dynamic workflows of both MPI- and non-MPI-driven codes.

\subsection{A Scalability Study: 100 Million Fusion Simulations on Sierra}
\label{JAG-100M}
The primary goal of \merlin~is the facilitation of large-scale simulation
ensembles on HPC systems.  This section tests \merlin's ability to scale
by showing how it created an unprecedentedly large fusion
simulation dataset, consisting of the multi-variate results of
roughly 100 million individual simulations on the Sierra Supercomputer at
Lawrence Livermore National Laboratory. A subset of the data~\cite{rushil:jag_data} has been released as an open-source dataset for the community-driven exploration of
ML models suited specifically to scientific applications.

The aim of inertial confinement fusion (ICF) is to compress a hollow shell of cryogenic 
deuterium-tritium (DT) fuel to thermonuclear conditions. The spherical DT ice shell is 
encased in an ablator material, the outer surface of which is heated indirectly via x-rays 
generated in an encasing hohlraum (in another variant, the laser directly heats the 
capsule). As the capsule surface ablates, the spherical shell implodes and 
compresses the enclosed fuel to high temperature and density. The goal is to have the gas at the center 
of the shell ignite a fusion burn wave, which consumes the DT shell and releases large 
amounts of fusion energy. Achieving fusion ignition in the laboratory is one of the key 
goals of the National Ignition Facility (NIF) and would mark a major step towards the ultimate goal of clean energy from fusion-based power plants.

JAG~\cite{gaffney2015data} is a semi-analytic
model of ICF implosions in 3D. As initial conditions, it takes two 0-dimensional
physics variables and three 3-dimensional capsule perturbations and evolves
an ICF capsule through the final stages of a NIF experiment. In the process
it produces scalar, time-series and hyper-spectral ray-traced images of the
implosion, which can be directly compared to experiments. Each simulation
runs in python on a single thread for approximately five minutes and outputs
48 images (4 frequencies $\times$ 3 viewing angles $\times$ 4 times), 16
time-series, 23 physics scalar quantities, 10 performance/system
scalars, plus the simulation input parameters and associated meta-data.

This study used \merlin~to parallelize the execution of JAG during early access time
on the Sierra supercomputer. The goal was to create a large collection of simulation results, which could be used to explore the relationships between ICF observables and to train ML models for analysis of NIF experiments (see Ref.~\cite{Anirudh9741} for the ML research using these data). Since ML model fidelity generally improves with the data quantity, the essential aim of the study was to run JAG as many times as possible and automatically pre-process and package data for subsequent analysis. Although JAG itself does not make use of
Sierra's GPUs, both Sierra's scale (no. of CPUs, size and speed of parallel
filesystem) and the desire to have the
dataset present on the system for GPU-driven ML applications served
as motivation for using Sierra.

\begin{figure}[htbp]
\begin{center}
\includegraphics[width=\columnwidth]{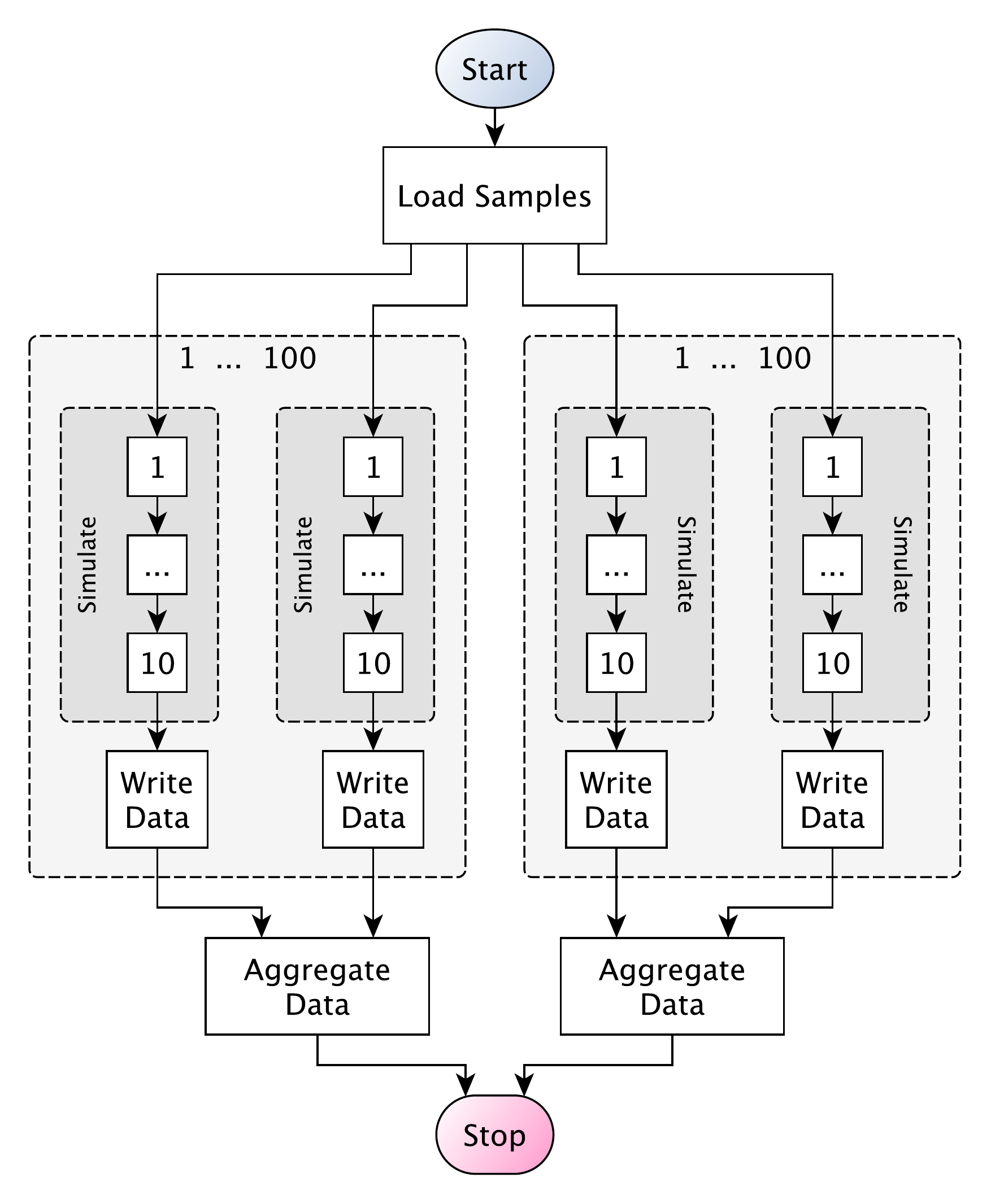}
\caption{Hierarchical JAG workflow the leverages tasks of simulation bundles to scale production. Each task bundle consists of 10 simulations processed in serial whose results are collected in memory and then dumped to disk. Every 100 tasks, the bundled results are aggregated into files of 1000 simulations apiece.}
\label{fig:jag_diagram}
\end{center}
\end{figure}

To maximize scalability, this study deployed a workflow (Fig.~\ref{fig:jag_diagram}) that combines individual simulations
into hierarchical meta-task ``bundles" of tasks and data. Each bundle consisted of 10 simulations which
were dumped via Conduit~\cite{conduit_2019} into hdf5 files, with 100 files per leaf
directory. Once each leaf directory was filled, an aggregation
task collected the bundled files into a single hdf5 data file containing 
1000 simulations. Each simulation produces approximately 300 kB of
zipped hdf5 data and a 1000-simulation aggregated bundle is roughly 300 MB.

The aim was to produce 100M simulations using stair blue
noise~\cite{Kailkhura:2016:SBN:2980179.2982435} sampling across 5 dimensions
precomputed and stored in 100 independent binary files, which were read
asynchronously during task creation.

The study used Flux~\cite{ahn2018flux} to launch Celery workers across multiple
independent batch jobs of varying sizes (64, 128, 256, 512, and 1024 nodes
of 40 workers each, one per core) and to dynamically allocate resources as workers requested them for a new simulation bundle. Each batch job submitted itself as a
dependent job via jsrun~\cite{jsrun} to create a ``worker farm" of batch requests. This
scheme allowed any holes in the scheduling system to be filled with a
combination of worker jobs, thereby maximizing utilization and throughput. At
peak, the scheme was able to run 1024 node and 512 node jobs simultaneously, such that 
61,440 concurrent workers were processing simulation requests,
communicating with the Rabbit server and writing to the disk. At this rate the system was running approximately 1 million simulations and writing about 100 GB
of physics data per wall-clock hour.

The initial run of tasks showed approximately a 70\% completion rate, with
the main reason for failing tasks being file system (I/O) and node failures
during the volatile early access period. A second run, in which \merlin~tasks
first crawled the directory tree and resubmitted missing simulations back to
the Rabbit task queue brought the success rate to 85\%. After one final pass of
task resubmissions, 99755022 simulations completed successfully, with
only 220978 failing due internal (physics) errors. The dataset totaled 24TB spread out across
99976 files. Its size (in number of samples and total volume) and complexity (scalars, 
time series, hyperspectral images) alone make it a cutting edge scientific dataset
(see Sec.~\ref{sec:intro}).

The study took advantage of some features of the \merlin~workflow system.
Firstly, the producer-consumer task
queuing system naturally separated what is to be run from where it is to be run, allowing for the exploitation of multiple concurrent batch jobs
of different allocation sizes. This increased system utilization and
reduced batch queue wait times. Secondly, the hierarchical data directory
system, the use of hdf5, and task result bundling allowed the
asynchronous creation of a large dataset without the need for file locking
or I/O coordination. Meta-tasks (that run multiple instances of JAG at a time) exploited on-node memory for the temporary
storage of simulation results within a bundle prior to writing to disk.
Thirdly, the integration with Flux for job launching
enabled the just-in-time launching of tens of thousands of simultaneous
simulations (at a peak rate of over 250 / second).
This allowed the task-queue system to coordinate the sequence of
work to be done without the explicit need for job and resource scheduling.
Finally, the use of independent atomistic tasks created workflow
resilience via a natural re-submission framework. Simulations that did not
complete were easily identified through 
exception handling during the run, or through detection of corrupt data after-the-fact, and
resubmitted to the task queue system. The decoupling of work from compute resources meant that workflow ``cleanup" could happen at a later time
with a smaller allocation.

\subsection{An Iterative Workflow Study: Optimizing the Design of a Fusion Experiment}
\label{sec:optimization}
The example in this section tests \merlin's ability to perform iterative simulation studies of MPI-driven codes. This workflow archetype, which consists of running a set of simulations, processing the results, performing analysis and then choosing a new set of simulations is a cornerstone of scientific computing. It encompasses such aims as simulation-driven design and optimization, active learning, and experimental analysis. In this section, we apply \merlin~to the machine-learning
enhanced optimization of a nuclear fusion experiment.

A major challenge of NIF experiments is that of manufacturability precision. In particular, the target capsule and laser system are designed and optimized together {\em in silico}, such that the requested laser pulse and requested capsule geometry are meant to be perfectly matched. However, due to finite experimental precision (for instance in the delivery of the laser power source or capsule manufacturability tolerances), the as-delivered laser pulse and capsule are not as intended, nor are they guaranteed to be optimally combined.

The study~\cite{peterson2020engineering} discussed in this section attempts to re-optimize the target capsule to align with the delivered laser pulse of a particular experiment and in so doing to take into consideration expected levels of capsule manufacturability variations. In other words, the goal is to maximize the expected nuclear yield under random draws from a prescribed capsule geometry. This problem setup necessitates constructing an optimization cost function that for each design point considers a number of small variations about that design point. Furthermore, the problem is one of constrained optimization that can consider expert judgment for ``valid" regions of parameter space: it is not sufficient to merely maximize yield, but rather to maximize yield subject to other simulation outputs (for instance, such that the capsule implosion velocity remain below a threshold, above which the experiment is unlikely to behave as predicted by simulations). Additionally, ICF capsule geometries can have 5-15 design parameters, which means that an iterative approach is likely to be more successful than a single-iteration approach (e.g. as done in Ref.~\cite{peterson-zonalflow}, which optimized a ML model built on 60,000 simulations in 9 dimensions). Another challenge is that the multiphysics design code, HYDRA, can be of moderate computational cost (if run in 1D), taking approximately 10-15 minutes on one core. This makes it too costly to run directly when calculating the objective function, but quick enough to where workflow overhead could become a significant fraction of compute time. Calculating the physics variables that feed into the objectives and constraints requires post-processing of simulation data (both raw data and intermediate data calculated {\em in situ}), which means that the vast majority of data produced is of little value for optimization, but might be of interest for subsequent analysis of any resulting design.

\begin{figure}[htbp]
\begin{center}
\includegraphics[width=\columnwidth]{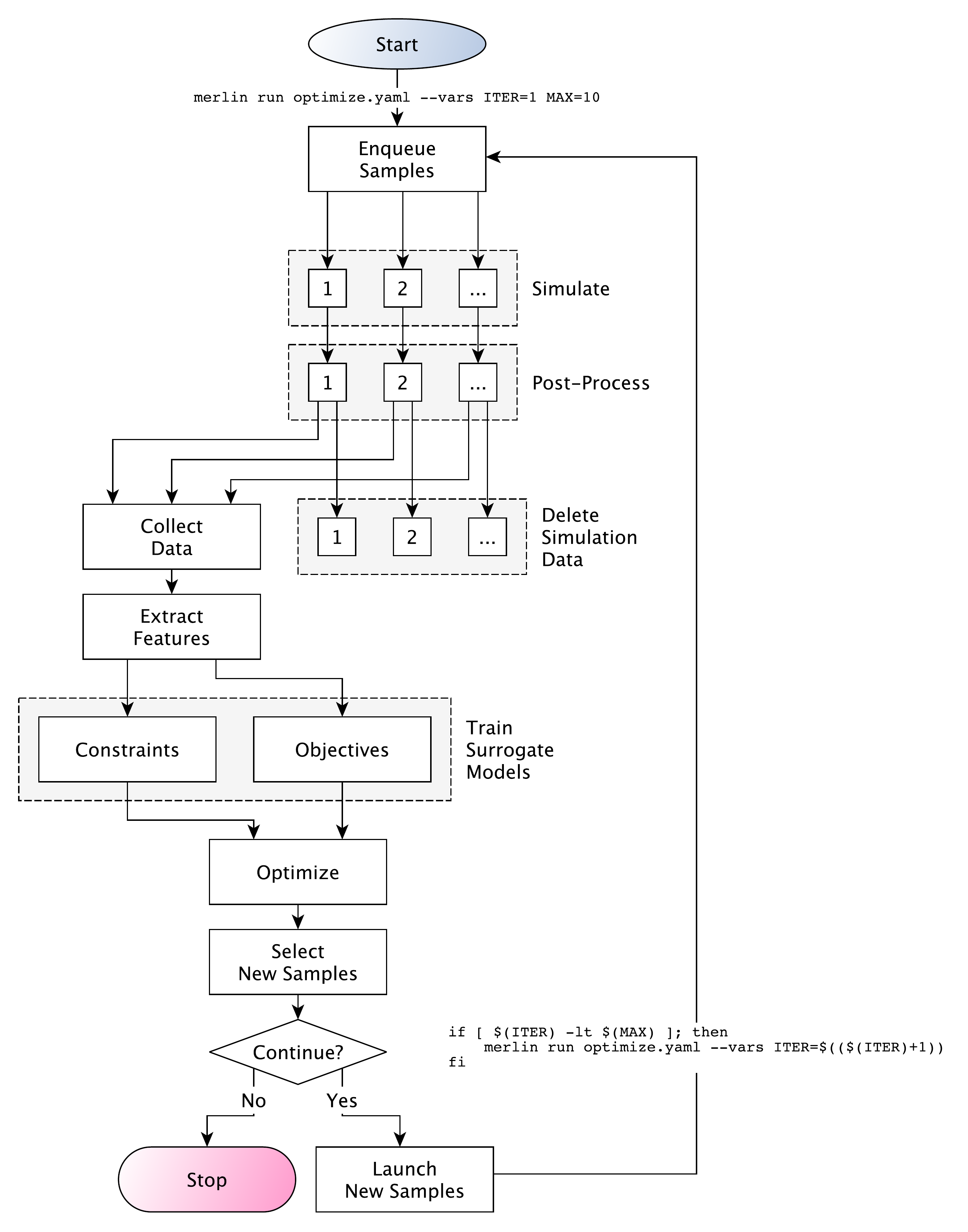}
\caption{Iterative workflow for surrogate-augmented optimization of fusion experiments. The optimization chain is initiated with a single call to \merlin~that sets the maximum number of allowed iterations. Subsequent iterations are automatically called with a worker task call that uses bash syntax to determine if another iteration should be launched.}
\label{fig:opt_diagram}
\end{center}
\end{figure}

Fig.~\ref{fig:opt_diagram} shows the workflow used to re-optimize the NIF ICF target 
capsules under uncertainties and constraints. The workflow leverages the producer-
consumer model to perform a number of iterative loops. During each iteration, workers 
running under Flux launch and post-process simulations asynchronously.
After a simulation is done post-processing, a worker deletes its raw data to save space and 
inode counts. Once all of 
the post-processing is completed, another task collects the data and extracts those features 
relevant for optimization. Then, a worker trains machine learning surrogate models for those 
features extracted from all previous iterations and saves them to disk. Subsequently, an 
optimization task loads those surrogate models and feeds them into an optimization 
algorithm that queries the models several times at each design point to produce both the 
cost function and evaluate the constraints. Once the surrogates have been optimized, a new 
search area is 
defined and new simulation samples chosen. Finally, the new 
samples are re-queued with a final worker call to \merlin~that also increments a variable 
defined in the workflow file to keep track of the iteration count.

The workflow was tested on the Lassen supercomputer, with 5 independent 2-node jobs of 
workers. Like the ``worker farm" concept of the JAG study, each job resubmitted a 
dependent child job.  With 80 total worker threads per job, the scheme allowed for up to 400 
concurrent workers when all jobs were running at the same time. However, at any given 
time any of the 5 of the batch jobs could be executing and contributing to the study. Each 
iteration produced 384 new simulations (128 around the best existing point, 128 at the 
predicted optimal and 128 connecting the two regions). To avoid wasting resources (for 
instance, by idling working while waiting for a single iteration to complete or while waiting for the last post-process task to complete prior to the synced collection step), multiple studies 
were launched at once under different constraint conditions, so that there was always work 
queued. This means that at any given time, workers on the same batch allocation were 
potentially working on different problems. Since both the re-queuing of simulation iterations 
and the resubmitting of workers to the batch system were automatic, the process was 
largely hands-off and produced roughly 100,000 simulations over the course of a few days 
calendar time (25,000 CPU-hours for the simulations spread out among multiple workers 
and batch allocations). A few optimization chains had to be manually resubmitted when 
system failures caused a particular chain to stop, much like during the 100M JAG study. But 
since the work is decoupled from the workers, this could be done without pausing or 
relaunching batch job allocations.

The optimization study benefited from a few features of \merlin. Firstly, the decoupling of 
producers and consumers meant that iterations could be automatic and computer resources 
used to work on multiple iterations and problems simultaneously without having to wait for 
dedicated batch jobs. It also allowed optimization chains that failed due to system issues to 
be resubmitted in real time. The shell-like API allowed for flow-control logic and the iterative 
and dynamic launching of chains, as well as the stitching of MPI-driven codes with
post-processing, optimization and machine learning tools, and the easy integration with Flux 
for on-demand resource allocation (e.g. combining multiple instances of HYDRA onto a 
single shared node).

\subsection{A Mixed Parameter-Sample Nested Workflow Study: Simulating COVID-19 Intervention Scenarios}
\label{sec:covid}
In this section, we describe a study that takes advantage of \merlin's low-overhead HPC-based producer-consumer
workflow model to create a cascading workflow. The goal of this work~\cite{anirudh2020accurate,thiagarajan2020machine} was to explore
possible non-pharmaceutical intervention scenarios to slow the advance of the COVID-19 pandemic with an agent-based epidemiological model, \epi~\cite{epicast}. \epi~is a large-scale parallel MPI-based model of infectious diseases originally built to model influenza. As an existing tool, \epi~was attractive for its potential use in the triage of possible intervention measures (e.g. whether to close schools or restaurants), but it (obviously) hadn't been applied to COVID-19 before. Doing so came with a number of challenges that \merlin~helped overcome.

\epi~uses the latest census data to create a virtual population of agents that mimic population characteristics, such as age distribution, household size, job type, commuting patters, etc., on a census tract level (roughly 2000 people), which are subsequently combined to approximate counties, metropolitan areas, states, and even the entire nation.
Unsurprisingly, the model depends on a number of parameters describing the initial conditions of an outbreak, disease biology, and social variables like the expected compliance with social distancing. Some of these parameters are ``global" in that they are similar in different locations (e.g. the innate infectivity of a particular disease), while some are ``local" in that they could be unique to a particular geographical region. In order to effectively use \epi~to either forecast potential outcomes or explore mitigation scenarios, all of these parameters must be first accurately calibrated.

Essentially, early in the outbreak, \epi~researchers needed to quickly find new valid parameter sets and then use these sets to launch more simulations of possible projection scenarios. With ground-conditions frequently changing and uncertain, these parameters were unknown. The research need therefore was the ability to run large quantities of small-scale simulations with low-overhead (\epi~itself could simulate a single metropolitan region under one scenario/parameter set on 64 cores of one Knights Landing (KNL) node in approximately 3 minutes) to find likely parameter sets valid for the disease and then to follow up each set with a series of simulations projecting forward in time and studying the effects of different possible intervention scenarios. Finally there was a need to organize and automate the distinction between ``global" and ``local" parameter sets and fold in post-processing and data packaging for later analysis. With the time pressures inherent in the outbreak, \epi~researchers could not spend large quantities of time learning new workflow APIs and wanted to be able to exploit large computational resources made available to the problem across multiple supercomputer sites.

\begin{figure}[htbp]
\begin{center}
\includegraphics[width=\columnwidth]{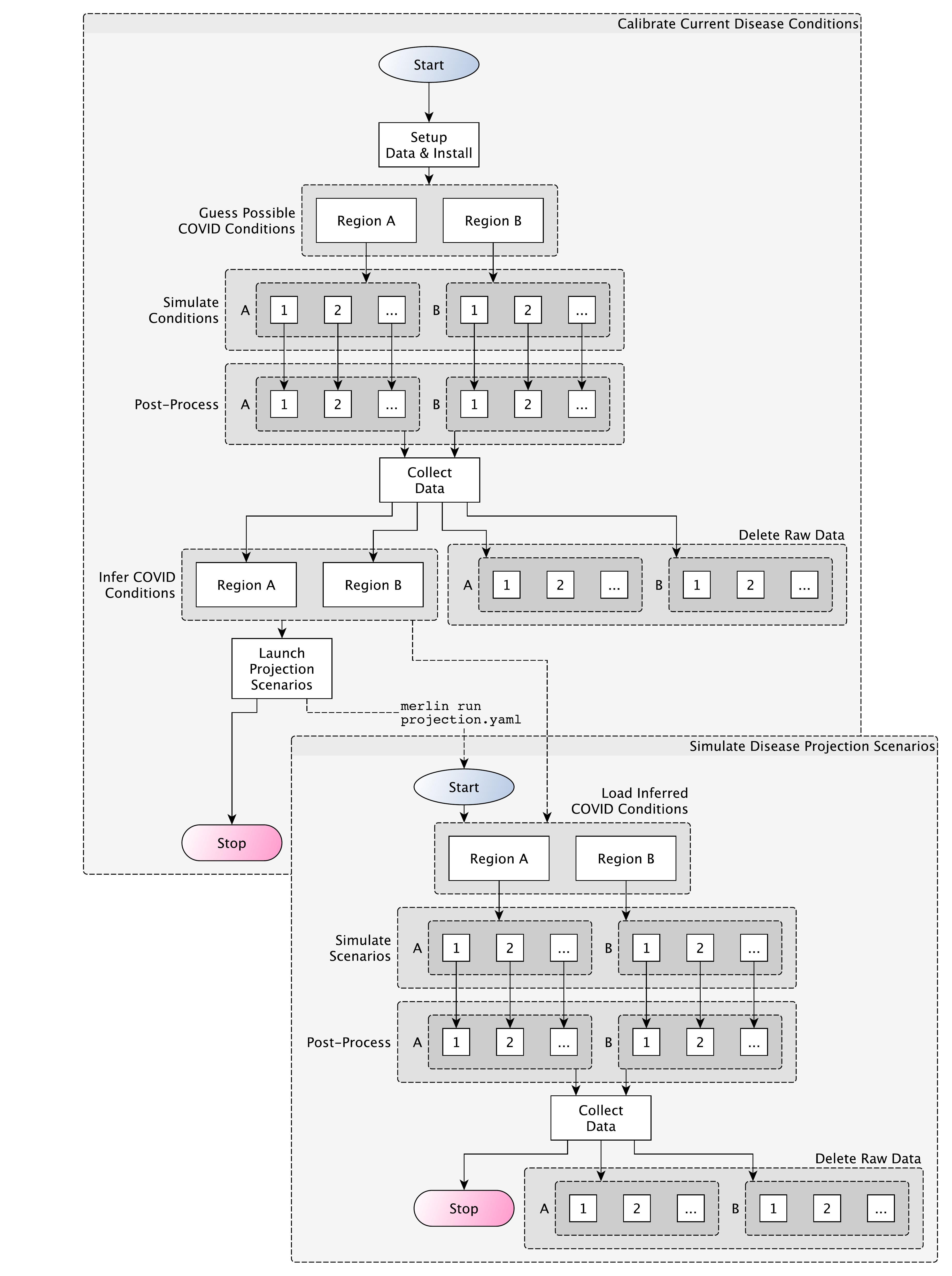}
\caption{Cascading workflow for studying the efficacy of non-pharmaceutical interventions for COVID-19. The first phase runs a series of agent-based epidemiological simulations to calibrate model parameters to fit current disease conditions and then launches the second phase, which simulates possible intervention scenarios.}
\label{fig:opt_diagram2}
\end{center}
\end{figure}

Figure~\ref{fig:opt_diagram2} shows a diagram of the two-step cascading workflow for the study of COVID-19 interventions. It defines DAG parameter sets per geographic region and reserves samples for disease conditions and intervention strategies. The first phase of the workflow (``calibration") pulls the latest COVID-19 case data and uses Flux to launch a series of ``pre-ensembles" to initialize the simulation code under different ground conditions and model parameters. At the end of the workflow, it refines its estimate of these parameters to match the data and then launches the second phase, which consists of longer running forecasting models under different intervention scenarios, followed by traditional data processing and packaging for subsequent analysis.

\merlin~provided a few enabling features for this study. Firstly, since worker steps can issue calls to {\texttt{merlin run}} and since this operation is very fast (see Fig.~\ref{fig:enqueuing_time_and_speed}), the two-step workflow could be automatic and dynamic. Secondly, the parameter-sample expansion distinction allowed for the asynchronous parallel calibration and study of multiple metropolitan areas. Thirdly the low workflow overhead per task meant that few resources were spent to workflow management, as compared to simulation compute. Fourth, the producer-consumer model meant that existing and surge-capacity workers could process many different scenarios without waiting for new batch allocations. Since the workers are decoupled from the work queue, multiple machines from the same compute center could be stitched together to form a federated system and increase throughput. Finally, by building upon existing open-source technology, \merlin~was able to be deployed at multiple supercomputing sites, allowing significant  computational resources to be allocated to the problem, while the shell-centered API allowed for the execution of the custom libraries needed to run the underlying code through Flux at each site.

\subsection{Conclusions from Real-World Examples}
This section presented three real world examples of \merlin's use for scientific studies. 
Together, some common key results and lessons emerge from \merlin's application.

Firstly, it's possible to use \merlin~to run large scale real world simulation ensemble 
studies. Good scalability from the JAG study came in large part to the ability to 
leverage Conduit/hdf5 for on-the-fly data bundling and aggregation. All studies used Flux to 
efficiently allocate resources when needed. The optimization study made judicious use of 
in-transit data deletion to save disk space. For these quickly running simulations, the 
relatively low overhead for a \merlin~task meant that compute resources could be devoted 
primarily to simulations. Another key to maximizing throughput in large-scale studies was 
the decoupling of producers and consumers, with separate batch allocations dedicated to 
running generic worker tasks that could pull work from shared queues. Not only did the
shell-based API allow for multiple codes to be stitched together into complicated 
workflows by 
non-experts, but it also allowed for the flexibility needed to run simulations at HPC centers: 
both \epi~and HYDRA required machine-specific special libraries to be linked and runtime 
environment variables set to properly run under Flux. The use of a shell-API combined with 
a producer-consumer model also allowed for workflows to break free of static DAG 
constructions to incorporate such constructs as looping and branching. Finally the 
distinction and separation between DAG-based parameters (with possibly complex 
dependencies) and samples (with simple dependencies but potentially large numbers) 
allows for both scalability in task creation and flexibility in workflow definition.

In two of the studies, system errors caused some jobs to fail, so the ability to resubmit 
tasks to the queue system without interrupting running workers helped minimize time spent 
waiting in a batch queue scheduler before correcting results, allowing resources to pick up 
naturally where the study left off.

The major challenge for studies using \merlin~in all cases was interfacing with third party 
libraries. In particular, although Flux and Conduit are not software dependencies, using 
them for their scalability benefits means that they must be installed and working separately 
at an HPC center. Furthermore, a Celery-supported queue server and backend (eg. RabbitMQ and Redis) also need to be deployed and instantiated 
on dedicated resources for workers to coordinate across batch allocations. Recent advances in 
container technology are making this installation easier, and such a strategy was used to 
deploy \merlin~at Lawrence Berkeley National Lab/NERSC and Oak Ridge National 
Laboratory for the COVID-19 study (using {\texttt {spin}}~\cite{spin} at NERSC and
{\texttt {slate/openshift}}~\cite{slate} at Oak Ridge). The biggest challenge to using \merlin~is 
the one-time up-front investment in the architecture supporting cloud-based technologies.
However, as compute centers move to incorporate more of these data services into their portfolios, porting~\merlin~to new locations should become easier. It may also be possible to explore containerized solutions on compute nodes within a batch allocation, establishing virtual pop-up servers, but creating a robust solution accessible to multiple nodes across batch allocations would be an open area of research.

\section{Conclusion}
\label{sec:conclusion}
\merlin~is a workflow framework designed to facilitate the creation of large
scale ensembles of HPC-driven simulations. With an eye toward future
integrated ML-driven workflows, it combines distributed asynchronous task-queuing software
with HPC technologies and allows for the integration with flexible hierarchical data formats and next-generation batch schedulers. 
\merlin's shell-like intuitive interface provides the flexibility required to
execute MPI-driven multi-physics simulations in a leadership-class HPC
environment.

In this paper we presented \merlin's motivation and design, tested its performance and scalability on an idealized workflow, and demonstrated its use on three real world scientific problems. The idealized test problem showed \merlin's capacity to scale workflow ensembles to several million samples with low overhead, empowered by \merlin's dynamic hierarchical task-generation algorithm.
Through the 100M JAG dataset, we demonstrated
how \merlin~can scale to thousands of nodes and coordinate tens of
thousands of asynchronous workers. When combined with Flux and
deployed on Sierra, it was able to process roughly 1 million ICF simulations per
hour to create a uniquely rich massive simulation ensemble dataset, which because of
its size and complexity, could serve as a benchmark
to stress advanced ML systems. The ICF optimization problem leveraged \merlin's consumer-producer model and low task queuing overhead to create a dynamic workflow that uses ML to steer the optimization of an MPI-driven code. Finally the \epi~COVID-19 study built a dynamic workflow that exploited the separation of parameters and samples that was both intuitive and scalable. Furthermore, the COVID-19 work was able to use \merlin's producer-consumer model to stitch together multiple HPC systems within a compute center into a coordinated resource, thereby enabling greater throughput of time-sensitive results. All examples benefited from surge computing, low task generation and execution overhead, and the shell-like API. Finally, all examples experienced random system failures that prevented some tasks from succeeding. Nonetheless, \merlin's provenance and resubmission framework allowed the users to recover from the failures in real time or after the fact, without having to wait for new dedicated compute resources.

\merlin~has been deployed at multiple HPC centers and is built using open-source components. Both \merlin\footnote{\merlin~is available under the MIT license: \texttt{https://github.com/LLNL/merlin}}~and a 10,000-sample subset of the JAG dataset\footnote{The JAG dataset is available via the LLNL Open Data Initiative: \texttt{https://data-science.llnl.gov/open-data-initiative}} are available publicly. The \texttt{merlin spellbook}\footnote{The \texttt{merlin spellbook} collection of workflow components is available at \texttt{https://github.com/LLNL/merlin-spellbook}.} is a companion project that contains useful routines and components for building workflows, such as sample generation, building ML surrogate models and data packaging.

Future work includes developing an API for periodic tasks and tools to facilitate job
monitoring, clean-up, and resubmission. Some examples come packaged with \merlin~(including the null example used to test performance), but additional templated workflows and worker
submission scripts could allow for increased user adoption. Finally it may be possible to write \merlin~interfaces to other workflow descriptions beyond the Maestro YAML file.

While \merlin~itself pushes the boundaries of contemporary simulation
ensemble creation, our work demonstrates above all that a confluence of
distributed and traditional HPC technologies represents a promising path
towards the realization of next-generation ML-integrated scientific computing.

\section*{Acknowledgment}
The authors wish to thank the Flux and Conduit developers, as well as the computing centers at LLNL, LBNL, and ORNL.

This work was performed under the auspices of the U.S. Department of Energy
by Lawrence Livermore National Laboratory under contract DE-AC52-07NA27344.
Lawrence Livermore National Security, LLC.
This document was prepared as an account of work sponsored by an
agency of the United States government. Neither the United States government nor
Lawrence Livermore National Security, LLC, nor any of their employees makes any
warranty, expressed or implied, or assumes any legal liability or responsibility for the
accuracy, completeness, or usefulness of any information, apparatus, product, or process
disclosed, or represents that its use would not infringe privately owned rights. Reference
herein to any specific commercial product, process, or service by trade name, trademark,
manufacturer, or otherwise does not necessarily constitute or imply its endorsement,
recommendation, or favoring by the United States government or Lawrence Livermore
National Security, LLC. The views and opinions of authors expressed herein do not necessarily
state or reflect those of the United States government or Lawrence Livermore National
Security, LLC, and shall not be used for advertising or product endorsement purposes. 

Released as \texttt{LLNL-JRNL-821884}.

\balance
\bibliographystyle{elsarticle-num}
\bibliography{merlin}

\end{document}